# The Role of Carbon Precursor on Boron Carbide Synthesis by Laser-CVD

A.J. Silvestre[1], M.J. Santos[2] and O. Conde[2]

[1]Instituto Superior de Transportes, R. Castilho, nº 3, 1269-074 Lisboa, Portugal

[2]Dep. de Física, Faculdade de Ciências da Universidade de Lisboa, 1749-016 Lisboa, Portugal

**Abstract.** This paper focuses on the synthesis of rhombohedral $B_4C$ (r-$B_4C$) coatings by $CO_2$ laser-assisted chemical vapour deposition (LCVD), using a dynamic reactive atmosphere of $BCl_3$, $H_2$ and $CH_4$ or $C_2H_4$. The influence of the carbon precursor on the deposition kinetics is discussed. The use of ethylene as carbon precursor presents several advantages over the use of methane, which is the conventional carbon precursor in CVD processes. These advantages are mainly related to its high absorption coefficient at the laser wavelength and a higher sticking coefficient, which enables to attain higher deposition rates and film thickness control at lower carbon precursor concentration. Films with carbon content from 15 to 22 at.% were grown at a deposition rate as high as 0.12 $\mu m.s^{-1}$.

## Introduction

Rhombohedral boron carbide (r-$B_4C$) is a highly refractory material that is of great interest because of its mechanical, thermal and electronic properties [1]. The r-$B_4C$ has been used as a neutron absorbent material in nuclear industry due to its high neutron capture cross-section [2]. Also of particular importance are the low specific weight and high hardness, the latter even surpassing diamond and boron nitride at temperatures over 1100 ºC [3]. Moreover, it presents a high melting point and modulus of elasticity, and has great resistance to chemical agents. This combination of properties makes boron carbide a prominent corrosion-resistant ceramic material for thin film applications. Furthermore, considering its high-temperature stability, large Seebeck coefficient and low thermal conductivity, boron carbide could find potential use as very high-temperature thermoelectric material for energy converters [1,4].

In previous studies conducted by our group, laser-assisted chemical vapour deposition (LCVD) of boron-carbon films was successfully achieved from a dynamic reactive atmosphere of $BCl_3$, $CH_4$, and $H_2$ at working pressures of 133 mbar [5-7] and 1 bar [8] using a continuous wave $CO_2$ laser. By comparing with conventional chemical and physical processes, enhancement of the deposition rate by one to four orders of magnitude was observed in the LCVD synthesis of boron carbide. In this deposition process, the laser beam impinges on the substrate at perpendicular incidence and the chemical reaction is thermally driven through heating of the substrate surface by the laser radiation. By scanning the composition of the reactive gas mixture, different phases could be deposited. In particular, the deposition of r-$B_4C$ within a wide chemical composition range was attained.

This paper aims to study the role of carbon precursor on the boron carbide synthesis by LCVD. Results of a comparative analysis of r-$B_4C$ films deposited by using the same boron precursor but changing the carbon precursor from $CH_4$ to $C_2H_4$ are presented. Because ethylene has a much higher sticking coefficient than methane [9], a higher deposition rate is expected. Moreover, ethylene, as well as boron trichloride, absorb the infrared $CO_2$ laser radiation through vibrational molecular excitation [10-12], whence direct heating of the gas phase by the laser beam is also expected, leading to a significant increase of the amount of deposited material.

## Experimental procedure

The experimental set-up used for deposition of r-$B_4C$ coatings was previously described in ref. [5]. Boron carbide films were deposited on silica substrates using a $CO_2$ laser as heat source, operated in cw $TEM_{00}$ mode at a wavelength of 10.6 μm, and a dynamic reactive gas mixture of $BCl_3$, $C_2H_4$ and $H_2$. Argon was used as buffer gas. The laser beam reaches the substrate at perpendicular incidence with a diameter of 12 mm. No focus lens was used since silica absorbs 84% of the laser radiation. Prior to their insertion in the reactor, the substrates were cleaned in ultrasonic baths of acetone and ethanol.

Since the substrates were always kept stationary under the laser beam, the experimental variables of the set-up are the laser output power (P), interaction time ($t_{int}$), total pressure ($P_t$) and partial flow rates of each gas ($\Phi_i$). In this study, the total pressure and argon flux were kept constant at 133 mbar ($1.33\times10^4$ Pa) and 400 sccm, respectively. The other experimental parameters were varied as shown in table 1. The relative amount of carbon and boron precursors in the reactive gas phase is characterised by the parameter $\varphi = 2\Phi_{C2H4}/(2\Phi_{C2H4} + \Phi_{BCl3})$ which took values between 0.05 and 0.13 in this study.

Table 1 - Process parameters for LCVD of r-$B_4C$ films using $C_2H_4$ or $CH_4$

| *Experimental parameters* | *Range of values* ($C_2H_4$) | *Range of values* ($CH_4$) |
|---|---|---|
| Laser output power [W] | 180 - 250 | 125 - 175 |
| Interaction time [s] | 90 | 30 - 90 |
| Partial flux of $BCl_3$ [sccm] | 41 | 34 - 40 |
| Partial flux of hydrocarbon [sccm] | 1 - 3 | 7.2 - 14 |
| Partial flux of $H_2$ [sccm] | 150 - 200 | 200 |

The r-$B_4C$ films deposited from ethylene were compared with those produced from methane [5-7], for which $0.15 \leq \varphi \leq 0.29$, the total pressure and argon flux were kept constant at 133 mbar and 430 sccm, respectively, the other experimental parameters being also presented in table 1.

The structure of the as-deposited films was extensively studied by X-ray diffraction at glancing incidence (GIXRD) of 1° with Cu-Kα radiation and the chemical composition was investigated by electron probe microanalysis (EPMA). The surface microstructure of the films was examined by scanning electron microscopy (SEM) and thickness profiles were determined by optical profilometry.

**Results and discussion**

**Surface morphology and microstructure.** Following the Gaussian laser energy distribution, all the deposited coatings have nearly circular shape. They are homogeneous, and exhibit the characteristic light grey shining colour of r-$B_4C$ and good adherence, like the ones previously produced from methane [5,7]. Also the microstructure of the films does not show any apparent dependence on the carbon precursor. SEM analysis of the surface of both films showed that their microstructure is nodular with a fine and uniform grain structure, with a mean grain size of few tenths of micron. However, the topography of the deposited material reveals a remarkable influence of the carbon precursor. When methane is used, Gaussian and top-flat profiles are only obtained at low laser power values. An increase of the laser power above 150 W and/or $t_{int} \geq 60$ s leads to a central depression associated with a lack of film material, yielding coatings with vulcano-like profiles. Replacing methane by ethylene in the reactive gas phase, the r-$B_4C$ films present thickness profiles always Gaussian or near flat, even at laser powers of 240 W and for $t_{int}$ = 90 s (Fig. 1).

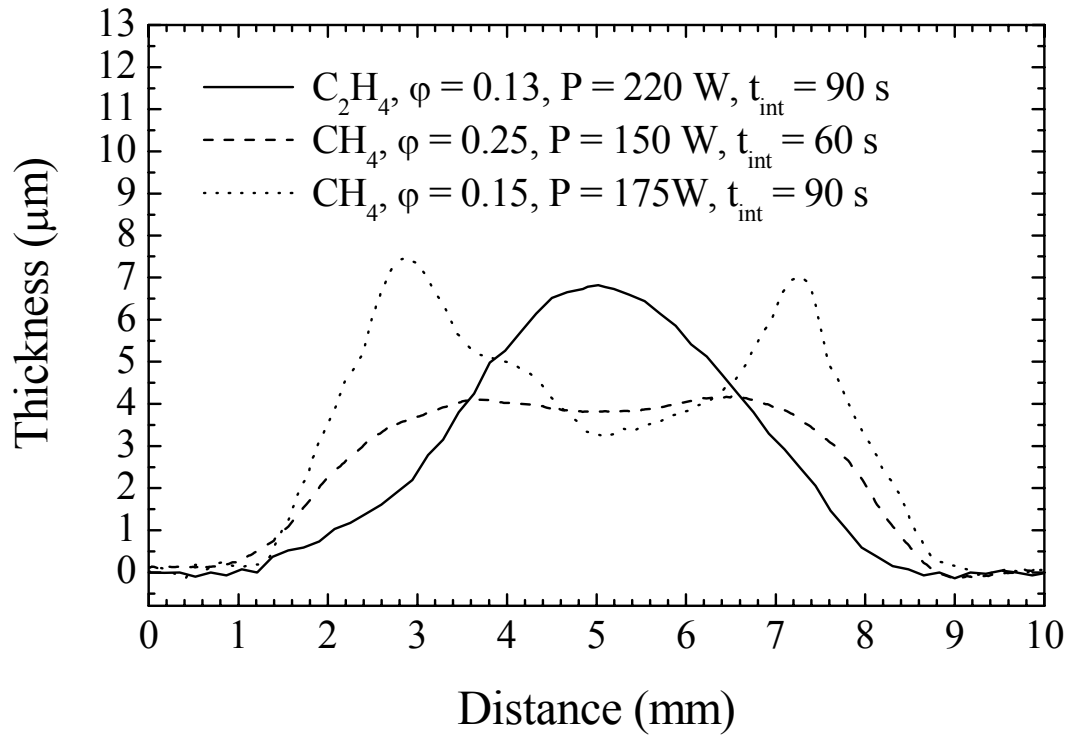

Fig. 1. Thickness profiles of r-$B_4C$ films deposited using $CH_4$ and $C_2H_4$ as carbon precursors.

**Chemical and structural analysis.** The EPMA chemical analysis of the films obtained with $C_2H_4$ allows to conclude that pure boron carbide was deposited all over the samples. Films with uniform composition along the spot radius were produced in a broad range of 15 to about 22 at.% C in agreement with the B-C phase diagram [13]. The X-ray spectra display the r-$B_4C$ diffraction pattern, showing narrow diffraction lines. Depending on the carbon content, the spectra match the JCPDS cards 33-0225 or 35-0798, corresponding, respectively, to the $B_{13}C_2$ and $B_{12}C_3$ stoichiometries. Moreover, the deposition of graphite was never observed and boron carbide phase was the only to be detected by GIXRD. Unlike the deposition from methane, for which the synthesis of r-$B_4C$ films with carbon content close to 20 at.% ($B_{12}C_3$ stoichiometry) is difficult to obtain without co-deposition of graphite [5,7], the present results indicate that pure r-$B_4C$ films with composition near the carbon-rich limit can be prepared by LCVD using ethylene as carbon precursor.

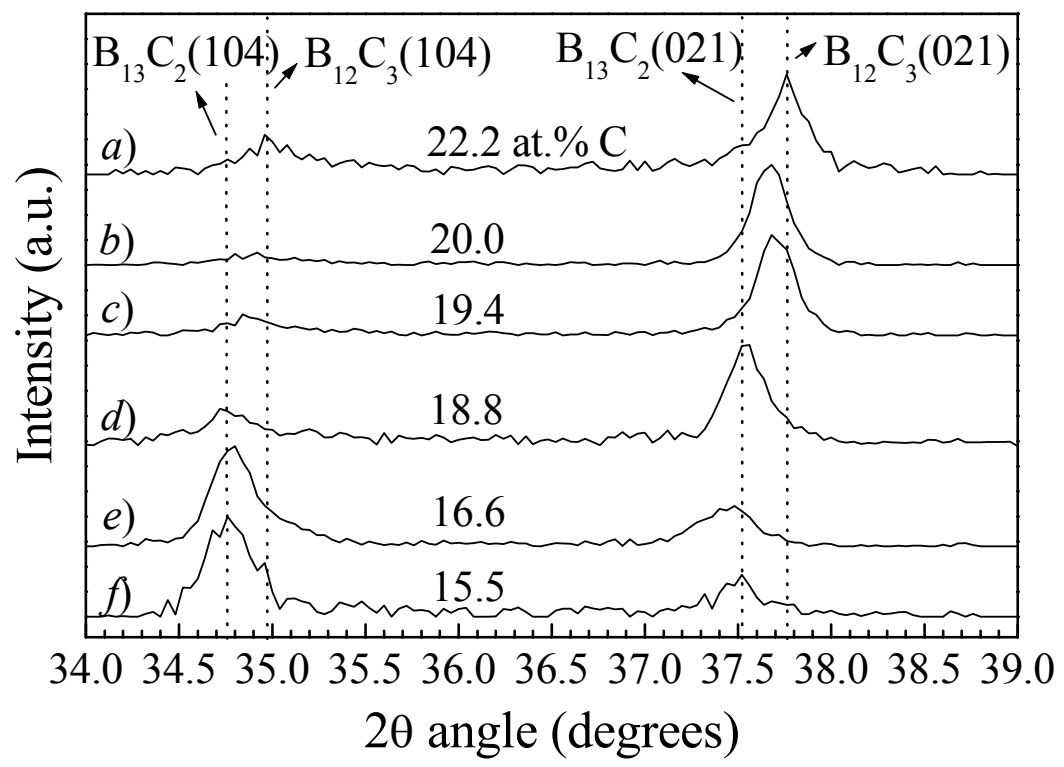

Fig. 2. GIXRD spectra of r-$B_4C$ films with different carbon contents, as measured by EPMA, prepared with $C_2H_4$ from the following experimental parameters: a) $\varphi$=0.13, $\Phi_{H2}$=200 sccm, P=220 W; b) $\varphi$=0.13, $\Phi_{H2}$=150 sccm, P=220 W; c) $\varphi$=0.09, $\Phi_{H2}$=200 sccm, P=240 W; d) $\varphi$=0.09, $\Phi_{H2}$=150 sccm, P=200 W; e) $\varphi$=0.05, $\Phi_{H2}$=150 sccm, P=240 W; f) $\varphi$=0.09, $\Phi_{H2}$=200 sccm, P=200 W.

The two major diffraction peaks characteristic of r-$B_4C$ compounds correspond to the (104) and (021) reflections. Their angular position and relative intensities vary with the carbon content of the deposited films, as can be seen in Fig. 2. The position of both peaks shifts left approximately 0.2° when the carbon concentration decreases from about 22 to about 15 at.% C. This observed shift is mainly related with the substitution of a carbon atom by a boron atom in the central C-B-C intericosahedral chain, leading to a C-B-B chain in the main diagonal of the rhombohedral structure of boron carbide [7].

A linear relationship between the $d_{021}$ values and the carbon content of the films was established. The least squares fitting of the data is shown in Fig. 3 and leads to the following result: at.% C = 758.4 - 309.7 $d_{021}$, in good agreement with previous LCVD results obtained from methane [14]. It should be noted, however, that this relationship is substantially different from the one reported by Jansson et al. [15] for CVD of boron carbide which was obtained considering samples with carbon content as low as 7 at.%, out of the homogeneity range of solubility generally accepted for rhombohedral boron carbide [13].

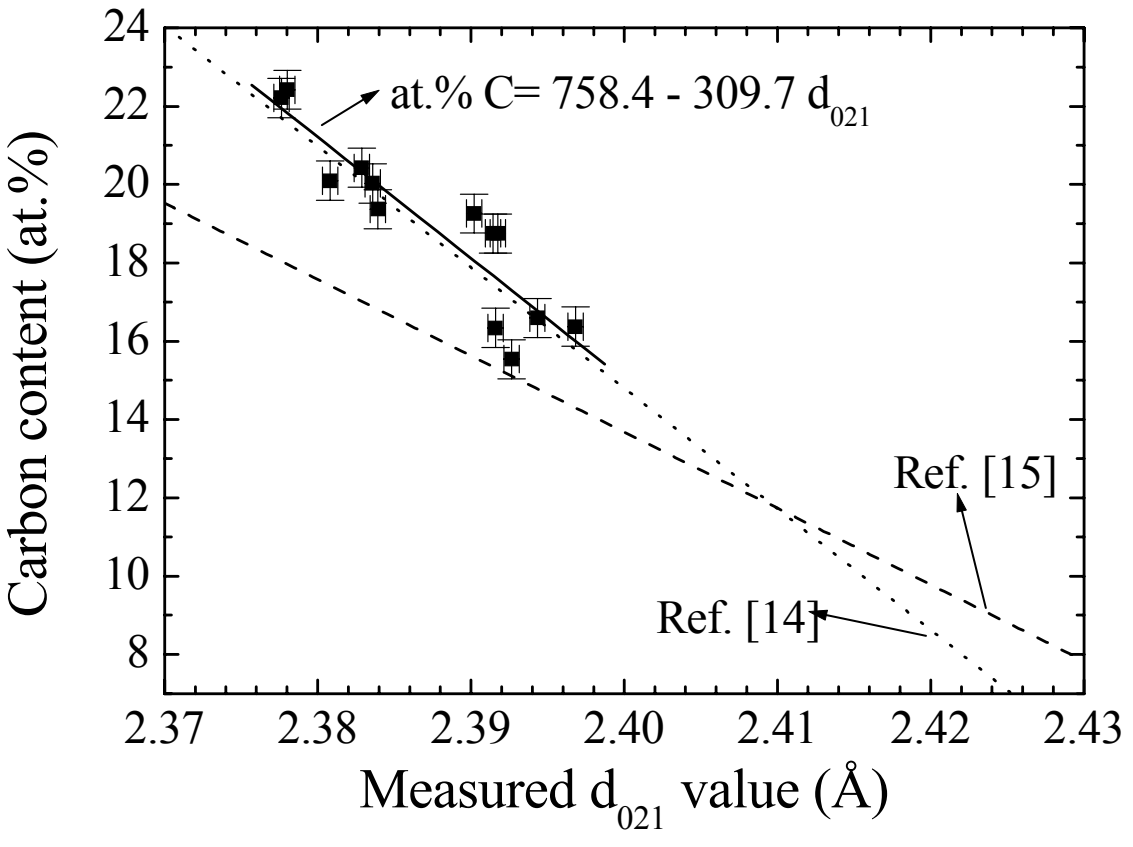

Fig. 3. Carbon content, as measured by EPMA, of r-$B_4C$ samples deposited from $C_2H_4$ *vs*. the interplanar distance $d_{021}$ measured by GIXRD.

Similarly to the films synthesised with methane [6,7], also those prepared from ethylene present the same inversion of the relative intensities of the (104) and (021) lines for carbon content lower than 17 at.% C, suggesting the development of a (104) texture (Fig. 2, spectra e and f). Although this trend to develop the (104) texture is not well understood, it is clearly independent of carbon precursor. Furthermore, preferential crystallographic orientations in this range of carbon contents have also been observed in other CVD processes, namely for low pressure CVD [16].

**Deposition rate.** The thickness distribution profiles provided a means to calculate the apparent deposition rate in terms of the maximum height, above the substrate surface, per unit of time. Following the calculation technique described in ref. [6], the surface temperature achieved at the centre of the films during deposition was estimated between 800 K and 1100 K for the films produced with methane, and between 780 K and 1430 K for those produced with ethylene. Fig. 4 shows the apparent deposition rate of the films deposited from both carbon precursors, as a function of the deposition temperatures referred to above. As can be seen herein, the deposition rates are strongly determined by the carbon precursor. For methane, as temperature increases, growth rates slow down and even decrease at the higher temperatures attained. The use of ethylene clearly favours the growth of boron carbide, enabling to double the deposition rate values of methane films for temperatures higher than 1000 K. Deposition rates as high as 0.12 $\mu m.s^{-1}$ were measured, one order of magnitude larger than those obtained in CVD processes [17,18]. It is also important to refer that these deposition rates were obtained with φ values lower than the optimised φ values for the r-$B_4C$ deposition from methane [6]. Three combined factors may explain the observed increase in the amount of deposited material when ethylene is used:

*i)* ethylene has a much higher sticking coefficient [9], thus a higher adsorption probability at the substrate surface, which favours the surface reaction;

*ii)* the strong absorption of the laser radiation by ethylene, namely the infrared double resonance on $\nu_7$ vibrational transition mode of the $C_2H_4$ molecule by the P(26) line of the $CO_2$ laser [11,12], also strongly contributes to promote the recombination of carbon and boron atoms to form boron carbide;

*iii)* for each carbon atom deposited from methane, four hydrogen atoms are released as products while in the case of ethylene only two hydrogen atoms are produced. The partial pressure of products in the deposition zone is then higher when methane is used, the inhibition of film growth occurs earlier and lower deposition rates are reached.

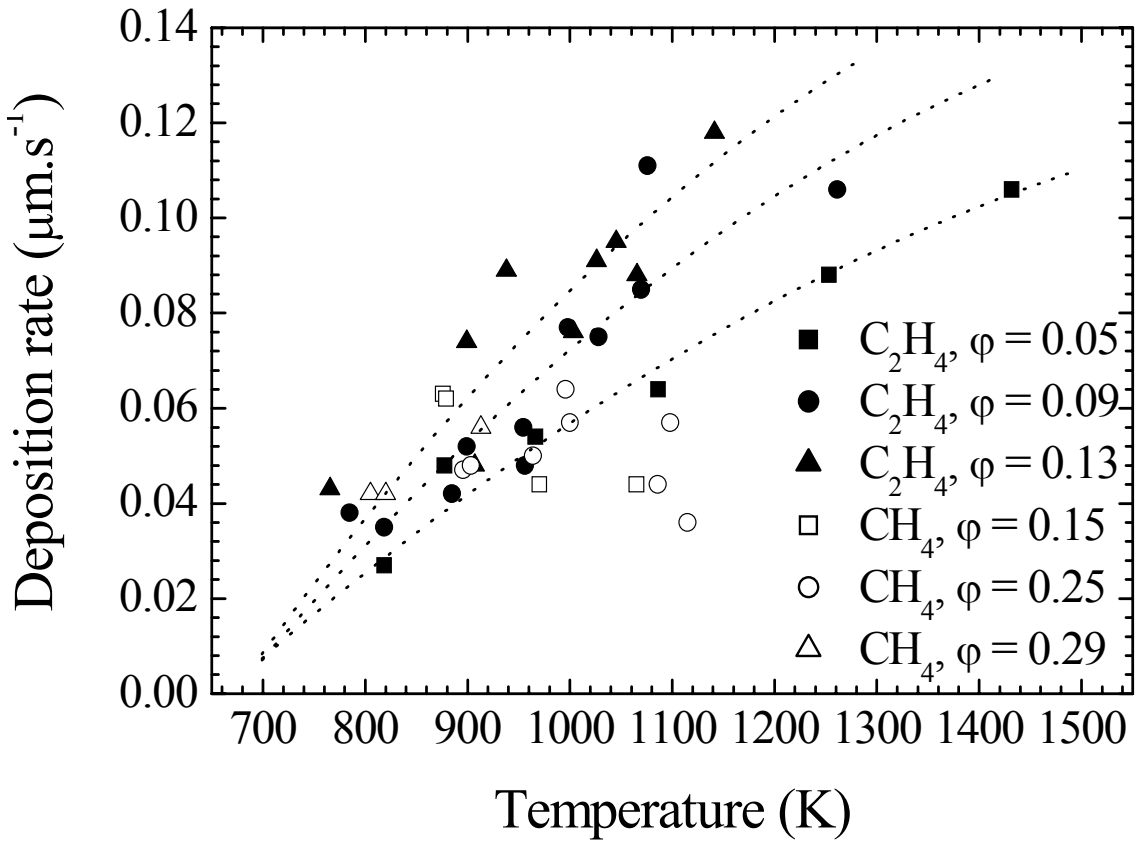

Fig. 4. Apparent deposition rate of $B_4C$ films prepared from different carbon precursors and carbon containing mixtures *vs*. estimated surface temperature at the centre of the films.

The wider range of deposition rate attained with $C_2H_4$ compared to that achieved with $CH_4$, suggests that LCVD of boron carbide from ethylene is much sensitive to the reaction temperatures induced by the laser beam, enabling to have a finer control of the deposition process and film thickness.

**Conclusions**

In this paper, it has been shown that ethylene is a suitable carbon precursor for the $CO_2$ laser-assisted CVD of rhombohedral boron carbide thin films. Films with good adherence, fine grain structure and carbon content from 15 to 22 at.% were grown at deposition rate as high as 0.12 $\mu m.s^{-1}$. Face to methane, which is the conventional carbon precursor in CVD processes, ethylene presents several advantages, namely those related with its high absorption at the $CO_2$ laser wavelength and its higher sticking coefficient, enabling to achieve higher deposition rates and film thickness control at lower carbon precursor concentrations. Moreover, this is an indication that the use of $C_2H_4$ seems to bring the LCVD process of r-$B_4C$ much closer to equilibrium conditions.

**Acknowledgements**

This work was partially funded by Fundação para a Ciência e Tecnologia, under POCTI program.